\documentclass{article}
\usepackage{float}

 \usepackage[preprint]{neurips_2026}

\usepackage{graphicx}
\usepackage[utf8]{inputenc} 
\usepackage[T1]{fontenc}    
\usepackage{hyperref}       
\usepackage{url}            
\usepackage{booktabs}       
\usepackage{amsfonts}       
\usepackage{nicefrac}       
\usepackage{microtype}      
\usepackage{xcolor}         
\usepackage{tcolorbox}
\tcbuselibrary{breakable}
\usepackage{amsmath}

\makeatletter
\renewcommand{\@noticestring}{Accepted in the 9th Workshop on Machine Learning and the Physical Sciences (ML4PS 2026).}
\makeatother

\title{The Great Chat in the Sky: 
Aligning Exotic Astronomical Data to Language with AstroBridge}

\author{%
\textbf{Ibrahim Habib}$^{1,2*}$ \quad \textbf{Kevin Li}$^{1*}$ \quad \textbf{Trung Nguyen}$^{1,3*}$ \quad \textbf{Gabriel A.~Patron}$^{1,4*}$ \\
\textbf{Gaurav Sekar}$^{1*}$ \quad \textbf{Nolan Koblischke}$^{1,5}$ \quad \textbf{Matthieu Le Lain}$^{1,6}$ \quad \textbf{Shashwat Sourav}$^{1,7}$ \\
\textbf{Aman Kumar}$^{1,8}$ \quad \textbf{Kartheik G.~Iyer}$^{1,9}$ \quad \textbf{Joshua S.~Speagle}$^{1,5}$ \\
\textbf{Itai Sfaradi}$^{1,10}$ \quad \textbf{Michael J.~Smith}$^{1,11}$ \quad \textbf{Ioana Ciucă}$^{1,12}$ \\[0.8em]
\normalfont\small
$^{1}$UniverseTBD \quad $^{2}$Suez Canal University \quad $^{3}$KU Leuven \\
$^{4}$University of Michigan, Ann Arbor \quad $^{5}$University of Toronto \\
$^{6}$Université Bretagne Sud \quad $^{7}$Washington University in St.\ Louis \\
$^{8}$IUCAA \quad $^{9}$University of Hawaii, Manoa \quad $^{10}$UC Berkeley \\
$^{11}$AstroAI, Harvard-Smithsonian Center for Astrophysics \quad $^{12}$Stanford University
}
\begin{document}

\maketitle

\begingroup
\renewcommand{\thefootnote}{}
\footnotetext{$^{*}$Joint first authors.}
\endgroup

\begin{abstract}
Astronomical observations include images, spectra, and light curves,
while the scientific literature describes these objects in natural
language. We present AstroBridge, a framework that connects these
observations to an LLM through a literature-assisted
caption-generation pipeline. We used catalog metadata to match
observations with scientific texts that describe the same objects.
A vision-language model generates descriptive captions from
representations of observations, matched passages, and available
metadata, creating observation-caption pairs without manual captioning.
We train AstroBridge to predict
captions directly from raw observations. Modality-specific frozen
encoders extract representations that a shared, trainable fusion
network aligns with the embedding space of an LLM. In caption-based
evaluation of spectra, AstroBridge achieves 91.5\% accuracy in
distinguishing quasars from galaxies, compared to 69.0\% from the base model prior to fine-tuning. Limitations remain in galaxy morphology, classification of minority supernova classes, and the factual quality of generated
captions. We believe that AstroBridge's framework is a step towards the goal of aligning the totality of astronomy's diverse, exotic data through the lingua franca of text.
\end{abstract}


\section{Introduction}
\label{sec:intro}

Scientific data is rarely in text form. Large
multimodal models (LMMs) address this gap by aligning modality-specific
encoders with large language models (LLMs), enabling scientists to
interact with their data in natural language. This approach has been
applied to clinical images~\citep{li_llava-med_2023}, amino acid
sequences~\citep{huo_multi-modal_2024}, and cardiac
signals~\citep{zhao_ecg-chat_2025}.

In astronomy, a single source can be observed through different
instruments, yielding images, spectra, and light curves.
General-purpose VLMs~\citep[e.g.,][]{comanici_gemini_2025}
process images and visualizations of other modalities, but lack
modality-specific encoders for the raw observations.
AstroLLaVA~\citep{zaman_astrollava_2025} and the Latent
Interpreter~\citep{kamai_talking_2026} connect images and spectra,
respectively, to LLMs, but remain limited to a single modality.
Other multimodal astronomical foundation models learn shared representations. For example, spectra have been aligned with text~\citep{vago_augmenting_2026}
and images with text~\citep{koblischke_2026}.
AstroCLIP aligns images
with spectra~\citep{parker_astroclip_2024}, while 
AION-1~\citep{parker_aion-1_2025} and
AstroPT~\citep{ref_smith2024,ref_siudek2026} connect multiple
astronomy datasets but do not natively provide a text interface for
interacting with the observations.

We present AstroBridge, a framework that combines a literature-assisted
data-generation pipeline with a shared natural-language interface,
allowing users---both human and agentic---to interact with astronomical
images, spectra, and light curves. We construct observation-caption
pairs without manual captioning and train AstroBridge to predict
these captions from raw observations. Modality-specific encoders
connect to the LLM through a shared, trainable fusion network.
The framework can be extended to new modalities by adding an encoder.
We will release the data pipeline, training code, and model weights
after deanonymization.

\begin{figure*}[htbp]
    \centering
    \includegraphics[width=\linewidth]{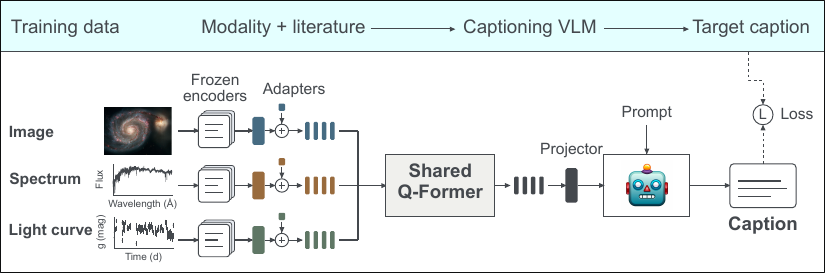}
    \caption{Overview of AstroBridge. An LLM generates training captions
from representations of images, spectra, and light curves,
matched literature, and catalog metadata. Raw observations pass through frozen modality-specific encoders, trainable projectors, and a shared
Q-Former with 64 learned queries. The resulting tokens are projected
into the LLM embedding space and prepended to the text input.
AstroBridge is trained to predict the generated captions from
 raw observations.}
    \label{fig:AstroBridge}
\end{figure*}

\section{Methodology}
\label{sec:methodology}
AstroBridge connects modality-specific encoders to an LLM through a
multimodal fusion network (Figure~\ref{fig:AstroBridge}). The complete caption-generation prompts, including
the literature and metadata provided for each modality, are given in
the Appendix~\ref{app:prompts}.

\subsection{Data collection and caption generation}

Using catalog metadata, we cross-match astronomical observations with
scientific text describing the same objects. We provide a VLM with the matched text and a representation of the
observation, supplemented where appropriate by catalog classifications
and physical measurements. We prompt the VLM to synthesize these inputs
into a descriptive caption, using the literature as background information and
keeping claims consistent with the supplied observations and metadata. AstroBridge is trained to predict these captions
from the raw observations.

\textbf{Images.} We cross-match the `Galaxy Mentions' corpus of literature
excerpts mentioning named astronomical
objects~\citep{galaxy_mentions_hats} against MMUv1.5
Legacy Survey DR10-south
imaging~\citep{dey_legacy_2019, ref_mmu2024}, using
HEALPix-partitioned spatial crossmatching via
LSDB~\citep{gorski2005healpix,lsdb_reference}. This yields 2,399 objects
with at least one literature mention, spanning 4,889 excerpts.
We convert \emph{g, r, z} flux cutouts to RGB composites using the survey
pipeline's arcsinh compression~\citep{Lupton_2004},
with two changes, namely, the affine stretch window is set per object from its
pixel percentiles, and white-balance weights are estimated from median
calibrated catalog flux ratios in our sample. Each object is rendered
twice, with a full-frame view and a tighter crop re-stretched using local
core pixel statistics. 

For each object, we prompt Gemini 3.7
Flash~\citep{google_gemini_3_7_flash_2026} to produce an image-only
description, a structured visual-property estimate, and a
literature-informed description. The latter uses crossmatched excerpts
to sharpen visual interpretation, with explicit instructions not to
assert causal physical scenarios or restate facts that are not
independently observable in the pixels. The fusion stage unifies the descriptions into the final training captions.

\textbf{Spectra.} We cross-match SDSS~\citep{sdss_dr17} and
DESI EDR~\citep{desi_edr} spectra from MMU~\citep{ref_mmu2024} with the
same corpus of literature used for images, using a matching radius of
$0.1^{\prime\prime}$. This yields 2,047 matched objects (1,845 from SDSS and 202 from DESI). We extract
catalog class and subclass where available, using SDSS and DESI DR1
classifications~\citep{sdss_dr17,desi_dr1}, together with emission-line
measurements for DESI spectra obtained using
FastSpecFit~\citep{fastspecfit}. 

We provide Gemini 3.8
Flash~\citep{google_gemini_3_8_flash_2026} with these physical
characteristics, the matched paper quotes, and a plot of the spectrum
to generate a descriptive caption. We instruct Gemini to use the plot
as the main source of information and the quotes only to enrich the
description where they do not contradict the plotted spectrum. 

\textbf{Light curves.} We train on 987 ZTF Bright Transient Survey
supernovae (563 SN Ia, 314 SN II, and 110 SN Ibc) with at least five valid
ALeRCE $g$- and $r$-band detections and matched
literature~\citep{perley_bts_2020,forster_alerce_2021}. Gemini 3.7 Flash
generates label-conditioned captions from the exact observations retained
by ATCAT~\citep{tung_atcat_2025}, using the literature only as background.
The evaluation uses 266 spectroscopically classified, object-disjoint Young Supernova Experiment (YSE)
transients (180 SN Ia, 71 SN II, and 15 SN Ibc), trimmed to their active
windows without altering retained measurements~\citep{aleo_young_2023}.

\subsection{Model architecture and training}

 Raw observations are passed separately through frozen AION-1
encoders for images and spectra~\citep{parker_aion-1_2025}
and ATCAT for light curves~\citep{tung_atcat_2025}. Per-modality projectors map their
outputs to a shared dimensional space, and a learned modality embedding
is added to each token. To merge these tokens, we use a shared
Q-Former~\citep{li_blip-2_2023}. A fixed set of 64 learned queries jointly
attends to the tokens of the present modalities, while attention
masks exclude absent modalities. The Q-Former produces a fixed-size
sequence of tokens, which is projected into the embedding space of
Qwen3.5-9B~\citep{qwen3_5_2026} and prepended to the text input.

We use a two-stage training strategy inspired by
LLaVA~\citep{liu_visual_2023}. In the
first stage, the encoders and LLM are frozen, and only the projectors,
the modality embeddings, and the Q-Former are trained. In the second stage,
we use Low-Rank Adaptation~\citep{hu_lora_2021} to fine-tune the LLM,
while the fusion network continues to train at a lower learning rate.
The encoders remain frozen throughout. We train on four NVIDIA GH200
Grace Hopper Superchips.

\section{Evaluation and Results}

We evaluate AstroBridge's ability to extract physical information from raw astronomical observations across the three modalities. We benchmark its performance against its unmodified base model and other VLMs that are provided the data in a textual or visual format.

\textbf{Spectra.} We prompt the models to caption each spectrum and use Gemini 4 Argon to answer scientific questions from the caption alone. We use the same evaluator for all models. We evaluated four downstream tasks: distinguishing quasars from galaxies, classifying source type (AGN, starforming, starburst, broadline), assigning a redshift bin, and extracting emission lines. For the first three tasks, the data are class-balanced, and we report accuracy. For emission-line extraction, we report the mean Jaccard index. As shown in Table~\ref{tab:results}(a), across all tasks, AstroBridge outperforms the base Qwen3.5-9B by a noticeable margin. The evaluated VLM baselines achieve lower scores than AstroBridge on these tasks, indicating the limitations of current small VLMs in extracting physical information from spectra. We leave further discussion of the results in Appendix~\ref{app:full-spectra-results}.


\begin{table*}[t]
\centering
\small
\caption{Performance on held-out test sets. AstroBridge directly processes raw observations via continuous encoders; in contrast, baselines require heuristic data representations, receiving inputs as rendered plots (plot) or raw numerical tables (table). $\pm$ values denote 95\% confidence margins from 10{,}000 bootstrap resamples. Bold = best, underline = second best.}
\label{tab:results}

\vspace{0.5em}
\noindent\textbf{(a) Spectra.} Performance on class-balanced downstream tasks. Except for emission lines, columns report accuracy.
\label{tab:spectra-results}

\vspace{0.3em}
\begin{tabular*}{\linewidth}{@{\extracolsep{\fill}}lcccc}
\toprule
Model &
\shortstack{QSO vs.\\Galaxy} &
\shortstack{Source\\type} &
\shortstack{Redshift\\class} &
\shortstack{Emission Lines\\(Jaccard)} \\
\midrule
Qwen 3.5 9B (plot)        & 69.0 $\pm$ 6.5 & \underline{41.5} $\pm$ 7.0 & 42.3 $\pm$ 5.7 & 28.7 $\pm$6.0 \\
Gemma 4 12B (plot)        & 61.5 $\pm$ 6.7 & 36.0 $\pm$ 6.5 & 35.3 $\pm$ 5.3 & 30.2 $\pm$ 6.0 \\
Qwen 3.8 27B (plot)       & \underline{73.0} $\pm$ 6.0 & 41.0 $\pm$ 6.8 & 35.7 $\pm$ 5.5 & 33.5 $\pm$6.1 \\
Muse Glimmer 30B (plot)   & 59.0 $\pm$ 7.0 & 38.5 $\pm$ 6.8 & \underline{45.7} $\pm$ 5.7 & \underline{34.1} $\pm$ 6.1 \\
Gemma 4 31B (plot)        & 51.0 $\pm$ 7.0 & 32.0 $\pm$ 6.2 & 40.0 $\pm$ 5.5 & 28.1 $\pm$ 5.9 \\
\textbf{AstroBridge}      & \textbf{91.5} $\pm$ 3.8 & \textbf{56.5} $\pm$ 7.0 & \textbf{90.3} $\pm$ 3.3 & \textbf{50.2} $\pm$ 5.7 \\
\bottomrule
\end{tabular*}

\vspace{0.8em}

\noindent\textbf{(b) Light curves.} Classification on the object-disjoint YSE test set.
\label{tab:lightcurve-results}

\vspace{0.3em}
\begin{tabular*}{\linewidth}{@{\extracolsep{\fill}}lccc}
\toprule
Model&
Accuracy (\%) &
Macro F1 (\%)&
Balanced acc. (\%) \\
\midrule
Qwen3.5-9B (table)
& 34.96 $\pm$ 5.27
& 25.80 & 32.05 \\

Qwen3.5-9B (plot)
& 58.65 $\pm$ 4.32
& 33.22 & 38.17 \\

Gemma 4 12B (table)
& 48.50 $\pm$ 5.64
& 36.29 & 38.29 \\

Gemma 4 12B (plot)
& 60.15 $\pm$ 3.76
& 34.84 & 36.59 \\

Gemma 4 31B (table)
& 69.17 $\pm$ 5.27
& \textbf{52.67} & \textbf{54.16} \\

Gemma 4 31B (plot)
& \underline{69.55} $\pm$ 4.89
& 43.59 & 44.78 \\



\textbf{AstroBridge}
& \textbf{77.07} $\pm$ 3.95
& \underline{52.25} & \underline{50.23} \\
\bottomrule
\end{tabular*}

\end{table*}

\textbf{Light curves.} We evaluate transient classification capabilities of the models on held-out YSE objects. Unlike in the other modalities, we directly ask the models to classify the light curves as SN Ia, SN II, and SN Ibc. We report accuracy, macro F1 and balanced accuracy in Table~\ref{tab:results}(b). As shown in Table~\ref{tab:results}(b), AstroBridge significantly outperforms its unmodified base LLM (Qwen 3.5 9B). AstroBridge achieves a macro F1 of 52.25\%, compared to just 33.22\% when the base model is provided with a rendered plot and 25.80\% when a raw numerical table is given. This may suggest that AstroBridge's encoders extract useful temporal features. Despite its 9B-parameter base, AstroBridge remains highly competitive with models more than three times its size, performing on par with Gemma 4 31B (52.67\% macro F1). We leave further analysis of these results in Appendix~\ref{app:full-lightcurve-results}.

\textbf{Images. }The image domain benefits from a variety of Large VLMs trained to bridge the two modalities. For this reason, we compare the vision performance of the base Qwen model, Qwen 3.5 9B, with our fine-tuned variant, AstroBridge, using 3 independent LVLMs as judges: Claude Sonnet 5 without thinking, Gemini Flash 3.8 with medium thinking, and GPT 5.6 Sol without thinking. We use Qwen 3.5 9B and AstroBridge to caption every image on the test partition under the same token budget. We use the Galaxy10 dataset ~\citep{leung2024galaxy10}. The assessment is conducted pairwise along two axes: description quality of image features and unsubstantiated claims. During the evaluation, the judge is shown images and captions to choose the winner along each axis, and a final request for the best overall caption is sent to the judge. The results in Table ~\ref{tab:pairwise-pct} highlight the judges' strong preference for AstroBridge's image captions. We leave further analysis of these results in Appendix~\ref{app:full-image-results}.

\begin{table}[t]
\centering
\caption{Pairwise comparison: AstroBridge vs. base Qwen 3.5 9B with equal token budget, judged blind
by three independent vision-language judges: Claude Sonnet 5, Gemini Flash 3.8, GPT 5.6 Sol. Thinking was off for both Sonnet and GPT, and Flash was used with medium thinking. $n=794$, with position-randomized A/B per object.
Percentages are of decided verdicts, higher meaning better.}
\label{tab:pairwise-pct}
\begin{tabular}{lccc}
\toprule
\textbf{VLM Judge} & \textbf{Overall} & \textbf{Better feature description} & \textbf{Fewer unsupported claims} \\
 & (\% AstroBridge wins) & (\% AstroBridge wins) & (\% AstroBridge wins) \\
\midrule
Sonnet 5 & 90.4\% & 90.4\% & 91.1\% \\
Flash 3.8 & 65.9\% & 64.2\% & 84.8\% \\
GPT 5.6 Sol & 90.6\% & 89.2\% & 93.5\% \\
\bottomrule
\end{tabular}
\end{table}

\section {Conclusions \& Future Work}
\label{sec:conclusion}

We introduce AstroBridge, a framework that connects multiple astronomical data modalities to an LLM. Due to the lack of human-annotated captions, we use a literature-assisted caption-generation pipeline to create the training data, which is used to align modality-specific encoders to the LLM through a shared fusion network. Because the pipeline relies on abundant unstructured observations and text, it is highly scalable and can be extended to new modalities by adding an encoder. Currently there is a difference in the version of model used for captioning images, compared with spectra and light curves which will be made uniform in a later revision. Our evaluation shows that our approach significantly enhances the LLM's ability to extract physical information from spectra and light curves, although it underperformed in galaxy morphology classification. 
Future work will study the effectiveness of passing multiple modalities simultaneously to improve physical understanding, such as combining images with spectra for redshift prediction or light curves with host-galaxy images for transients. 
Furthermore, we plan to scale our data-generation pipeline by incorporating a wider variety of textual styles over more modalities and larger data corpora.

\section{Acknowledgments}

This work used the DeltaAI advanced computing and data resource at the National Center for Supercomputing Applications (NCSA) through allocation PHY250358 (AstroLLaVA 2.0: Open Language Models for Multimodal Astronomy) from the Advanced Cyberinfrastructure Coordination Ecosystem: Services \& Support (ACCESS) program, which is supported by U.S. National Science Foundation grants \#2138259, \#2138286, \#2138307, \#2137603, and \#2138296. DeltaAI is supported by the National Science Foundation (award OAC 2320345) and the State of Illinois, and is a joint effort of the University of Illinois Urbana-Champaign and its National Center for Supercomputing Applications \citep{boerner_access_2023}.

We thank EleutherAI for hosting this project as part of their Summer Of AI Research (SOAR) program. We also thank them for compute and infrastructure support. We also thank Google for providing Google Cloud credits through the Gemini Academic Program Award, which supported this work.

\bibliographystyle{plain}
\bibliography{references}


\appendix
\newpage
\section{Prompts}
\label{app:prompts}

\subsection{Image caption generation}
\label{app:image-caption-prompts}

 \begin{tcolorbox}[
      colback=gray!5!white,
      colframe=gray!75!black,
      fonttitle=\bfseries\small,
      fontupper=\footnotesize,
      title=AstroBridge Captioning Prompt (Merged Blind Description + Property Estimate Stage),
      sharp corners,
      breakable,
      left=2mm, right=2mm, top=1.5mm, bottom=1.5mm
  ]
  \textbf{ABSOLUTE RULE -- THIS OVERRIDES EVERY OTHER INSTRUCTION BELOW:} You must NEVER output the object's name, any catalog designation (NGC/IC/UGC/M/Messier/PGC/2MASX/etc.), any host-galaxy label, or any other proper identifying string -- for this object, or for any OTHER object that happens to be mentioned in your input. Not partially, not as part of a longer phrase, not even if it would make the answer more complete or accurate. If you are ever unsure whether something counts as a name, leave it out. This matters more than completeness, more than detail, more than any other instruction here. A response that names anything has completely failed the task, no matter how good the rest of it is.

  You are an expert astronomical image analyst. You are shown one image of the target: the full field of view, a representative-color composite. These are representative-color composites from a wide-field optical imaging survey (channel mapping: z-band=red, r-band=green, g-band=blue).

  The target is the source nearest the center of the frame. If another source (a foreground star, a companion galaxy, a background object) appears elsewhere in the frame, describe it separately and clearly as a distinct source, not as part of the target. If two comparably bright components both sit near the center and it is genuinely ambiguous which is the target, say so explicitly rather than picking one.

  You will produce TWO independent pieces of analysis of the same image, Part A and Part B below. Treat them as two separate tasks -- do not let one task's framing, structure, or wording bleed into the other's; each is judged as if it were the only thing you were asked to produce.

  \textbf{PART A -- open-ended description:}\\
  Describe ONLY what is visible in the image itself.

  \textit{Guidelines:}
  \begin{itemize}
      \item Do not state facts you cannot determine from the pixels alone (no distance, redshift, age, or discovery history).
      \item Report only what this specific image shows. Naming a directly visible pattern (``a spiral pattern'', ``a smooth oval light distribution'', ``a point-like appearance'') is fine; inferring the class, physical state, or population that pattern implies is not -- that is background astrophysical knowledge, not observation.
      \item This is a stretched, non-linear display composite (not calibrated photometry). Describe only the apparent displayed color you see -- do not infer stellar age, star-formation activity, dust content, or population type from color alone.
      \item Use the displayed frame only for direction (``vertical'', ``horizontal'', ``toward the upper right'') -- never compass directions, since this composite's orientation on the sky is not confirmed to you.
      \item Describe morphology as a directly visible pattern, any structure actually visible (arms, bar, bulge, disk, knots, halo), texture, and any notable features. Inspect closely -- including faint, low-contrast, and off-center regions -- before concluding there is nothing more to report; report a marginal feature as uncertain rather than omitting it or overstating it.
      \item If the target shows any resolved structure at all, commit explicitly to three specific calls rather than leaving them unaddressed -- skip only the ones a target genuinely can't support. Each has three valid answers -- a specific pattern, its clear absence, or ``not clear either way'' -- and ``not clear'' is as complete an answer as either of the other two: (1) does the central region show a straight, bar-like feature through the middle, or clearly not, or is that not clear? (2) if spiral arms are visible, do they wind tightly with many close turns near the center, or loosely with few widely-spaced turns, or is that not clear? (3) if the object appears close to edge-on, does the central region bulge out noticeably thicker than the surrounding disk, or does the disk taper with no separate bulge, or is that not clear? Weave these into the flowing description -- do not answer as a separate checklist.
      \item If you see two or more components in close proximity, or a single envelope with a warped, irregular, or otherwise disturbed shape, make an explicit call: is there concrete visual evidence of an ongoing interaction or merger (a tidal tail, a bridge/filament of light, a double nucleus, a warped/asymmetric disk), or do the components look undisturbed and simply close in projection, or is that not clear? Use words like ``merging'' or ``interacting'' only when concrete distortion is actually visible -- undisturbed nearby sources are a ``close pair'', not a merger.
      \item If the target itself is compact, point-like, or shows no resolved internal structure, do not enumerate which specific structures are absent -- say plainly that it shows no additional resolved structure and stop there.
      \item Write as many sentences as the object actually warrants, typically 2--6. Do not pad length with speculative detail or classification language it doesn't need.
      \item Refer to it only with generic terms such as ``this galaxy'' or ``this object''.
  \end{itemize}

  \textbf{PART B -- structured property estimate:}\\
  Examine the target the way an observer would at the eyepiece or on a display -- with no catalog, photometry software, or measurements available to you. Only your visual judgment.

  First, classify the target's resolvedness, then follow only the matching branch below -- do not answer both:
  \begin{itemize}
      \item \textbf{UNRESOLVED OR MARGINALLY RESOLVED}: describe only multiplicity (any hint of more than one component, even faintly), light profile (sharply peaked vs.\ softly extended), elongation or asymmetry, any surrounding halo, apparent color, and any separate nearby source. Do NOT discuss spiral arms, a bar, rings, disk/bulge structure, star formation, or dust for a target in this branch.
      \item \textbf{CLEARLY EXTENDED}: give your best visual estimate of whichever of the following the image actually supports -- skip an item entirely if there is nothing to say about it: color balance (systematically redder/bluer, uniform or varying); extent and shape (circular vs.\ elongated, and direction); structure (bulge, disk, spiral arms, bar, rings, knots, irregular clumps, and their arrangement); clumpiness/obscuration-like structure (blue knots, patchiness, dark lanes -- as a pattern only, not as ``star formation'' or ``dust'').
  \end{itemize}

  \textit{Guidelines:}
  \begin{itemize}
      \item These are your own visual estimates, not measurements you were given -- state them as estimates (``appears to be...'', ``looks roughly circular...'') rather than precise facts.
      \item Do not state facts you cannot estimate from the image (no distance, redshift, age, or discovery history).
      \item Same no-priors, color, orientation, and inspect-closely rules as Part A.
      \item Write as many sentences as the object actually warrants -- often just 2--3 for an unresolved target, up to 5--6 for a clearly extended one.
      \item Refer to it only with generic terms such as ``this galaxy'' or ``this object''.
  \end{itemize}

  Return valid JSON only, no markdown fences, in exactly this shape:\\
  \texttt{\{"blind\_description": "<your Part A response>", "structured\_properties": "<your Part B response>"\}}

  Before you answer, check your own response one more time: does it contain any name, designation, or identifying string, for this object or any other? If so, remove it before responding. Check BOTH JSON fields for a leaked name, not just one.
  \end{tcolorbox}

\begin{tcolorbox}[
      colback=gray!5!white,
      colframe=gray!75!black,
      fonttitle=\bfseries\small,
      fontupper=\footnotesize,
      title=AstroBridge Captioning Prompt (Literature-Informed Description Stage),
      sharp corners,
      breakable,
      left=2mm, right=2mm, top=1.5mm, bottom=1.5mm
  ]
  \textbf{ABSOLUTE RULE -- THIS OVERRIDES EVERY OTHER INSTRUCTION BELOW:} You must NEVER output the object's name, any catalog designation (NGC/IC/UGC/M/Messier/PGC/2MASX/etc.), any host-galaxy label, or any other proper identifying string -- for this object, or for any OTHER object that happens to be mentioned in your input. Not partially, not as part of a longer phrase, not even if it would make the answer more complete or accurate. If you are ever unsure whether something counts as a name, leave it out. This matters more than completeness, more than detail, more than any other instruction here. A response that names anything has completely failed the task, no matter how good the rest of it is.

  You are an expert astronomical image analyst. You are shown one image of the target: the full field of view, a representative-color composite. These are representative-color composites (z-band=red, r-band=green, g-band=blue) of an object catalogued under the following name(s): \texttt{\{target\_names\}}. You are being told this name ONLY so you can tell which literature excerpts below are actually about this object versus some other object -- it must never appear in your response (see the absolute rule above).

  The target is the source nearest the center of the frame. If another source (a foreground star, a companion galaxy, a background object) appears elsewhere in the frame, describe it separately and clearly as a distinct source, not as part of the target. If two comparably bright components both sit near the center and it is genuinely ambiguous which is the target, say so explicitly rather than picking one.

  You are also given a small selection of excerpts from published scientific literature that mentions this object (selected for source diversity, not exhaustiveness -- a real description does not need to cite every paper ever written about a target). IMPORTANT: some excerpts discuss OTHER objects in the same sentence or alongside it -- comparison galaxies, field neighbors, companion sources. Any measurement, property, classification, or description attributed to an object other than \texttt{\{target\_names\}} does NOT describe the object in the image. You must identify and completely disregard that content; only reason from statements that are actually about \texttt{\{target\_names\}}.

  \textbf{Literature excerpts:}\\
  \texttt{\{literature\_block\}}

Using the image itself as the primary evidence, and the literature only to sharpen vocabulary and framing, write ONE expert-level description that:
\begin{itemize}
    \item Never infers or selects a causal physical scenario from the image. If the literature discusses a merger, interaction, lens, AGN, resonance, starburst, or a population state (young/evolved/quiescent/gas-rich/etc.), describe only the corresponding \textit{visible} feature (e.g.\ ``an asymmetric outer light distribution'', ``a faint arc-like feature'', ``a compact core'') and omit the proposed physical cause entirely. A visible feature being compatible with a scenario the literature raises is not confirmation of that scenario -- never report it as if it were.
    \item May use precise structural vocabulary from the literature (naming a visibly-present ring, bar, or arm using the literature's own term for it) only when the image itself clearly shows that structure -- never to describe a structure the image does not show.
    \item If the target itself is compact, point-like, or shows no resolved internal structure, do not enumerate which specific structures the literature discusses (spiral arms, a bar, rings, dust lanes) as absent -- say plainly that no additional resolved structure is visible and stop there, regardless of how much the literature has to say about this kind of object.
    \item Does not restate specific non-visual facts (no distances, redshifts, dates, masses, or other papers' measurements) -- use them only to inform how you interpret the pixels, not as facts to report.
    \item Ignores anything about identity, naming, discovery, or classification history -- for this object or any other.
    \item Reports only what this specific image shows (no unsupported inference of class/physical state/population from a directly visible pattern), describes only the apparent displayed color (never inferring stellar age, star formation, dust, or population type from it), and uses only frame-relative direction language (never compass directions).
    \item Is written as a single flowing paragraph of prose -- no headers, no bullet points, no markdown formatting -- with its length matched to how much the image actually shows. A visually simple or unresolved object should get a short paragraph (2--4 sentences), not an exhaustive structural breakdown just because literature context happens to be available.
\end{itemize}

Before you answer, check your own response one more time: does it contain any name, designation, or identifying string, for this object or any other? If so, remove it before responding. This is the highest-risk step in this entire pipeline because you were just shown the real name -- treat that as a reason for extra caution, not less.
\end{tcolorbox}

\begin{tcolorbox}[
    colback=gray!5!white,
    colframe=gray!75!black,
    fonttitle=\bfseries\small,
    fontupper=\footnotesize,
    title=AstroBridge Captioning Prompt (Fusion/Reconciliation Stage),
    sharp corners,
    breakable,
    left=2mm, right=2mm, top=1.5mm, bottom=1.5mm
]
\textbf{ABSOLUTE RULE -- THIS OVERRIDES EVERY OTHER INSTRUCTION BELOW:} You must NEVER output the object's name, any catalog designation (NGC/IC/UGC/M/Messier/PGC/2MASX/etc.), any host-galaxy label, or any other proper identifying string -- for this object, or for any OTHER object that happens to be mentioned in your input. Not partially, not as part of a longer phrase, not even if it would make the answer more complete or accurate. If you are ever unsure whether something counts as a name, leave it out. This matters more than completeness, more than detail, more than any other instruction here. A response that names anything has completely failed the task, no matter how good the rest of it is.

You are an expert editor producing ONE final caption for this astronomical object by reconciling three independent draft observations of it against the actual pixels. You are shown one image of the target: the full field of view, a representative-color composite. These are representative-color composites (z-band=red, r-band=green, g-band=blue) of an object catalogued under the following name(s): \texttt{\{target\_names\}}, told to you only for disambiguation -- it must never appear in your response.

\textbf{Draft 1} (unguided visual description):\\
\texttt{\{stage1\_text\}}

\textbf{Draft 2} (structured visual property estimate):\\
\texttt{\{stage2\_text\}}

\textbf{Draft 3} (literature-informed visual description):\\
\texttt{\{stage3\_text\}}

Treat all three drafts as proposals, not as ground truth -- independently re-examine the image yourself before deciding what to keep. Do not resolve a disagreement between drafts by preferring whichever one is more specific or more confidently worded -- a detailed, confident claim is not more likely to be correct than a hedged one, it may simply be a more specific hallucination. Resolve every disagreement by looking at the actual pixels: if a feature is visible to you, keep it regardless of how many drafts mentioned it (even if only one did); if a feature is not visible to you even though a draft claimed it, drop it or state it as uncertain.

A draft's SILENCE about a feature is not evidence against it. If one draft reports a specific, localized feature (e.g.\ ``two closely spaced peaks near the core'') and another draft simply didn't mention it, that is not two-against-one -- it is one concrete positive observation against zero contrary observations, since the other draft never claimed the feature was absent, it just didn't independently notice it. Only treat drafts as disagreeing when one makes a positive claim and another makes an explicit, specific contrary claim about the same region (not a generic ``no other structure visible''). When a specific positive claim from one draft survives your own look at the pixels as at least plausible, keep it -- stated with uncertainty (``may show...'', ``possibly...'') if you can't fully confirm it yourself, rather than silently dropping it because other drafts didn't independently mention it.

Write ONE final caption that:
\begin{itemize}
    \item If the target itself is unresolved or only marginally resolved, does not enumerate absent internal structures (spiral arms, a bar, rings, a disk, dust lanes) -- describe only what is actually visible (multiplicity, profile, elongation, asymmetry, halo, apparent color, separate nearby sources), matching whichever draft(s) correctly applied this rule rather than one that didn't.
    \item If the target shows any resolved structure at all, makes an explicit call on three specific questions rather than leaving them unaddressed -- agree with whichever draft(s) already made the call if your own look at the pixels supports it, or override them if it doesn't; skip only the ones the target genuinely can't support (a winding call needs visible spiral arms; a bulge call needs a close-to-edge-on target). Each has three valid answers -- a specific pattern, its clear absence, or ``not clear either way'' -- and ``not clear'' is as complete an answer as either of the other two, not a lesser one: (1) does the central region show a straight, bar-like feature through the middle, or clearly not, or is that not clear? (2) if spiral arms are visible, do they wind tightly with many close turns near the center, or loosely with few widely-spaced turns, or is that not clear? (3) if the object is close to edge-on, does the central region bulge out noticeably thicker than the surrounding disk, or does it taper with no separate bulge, or is that not clear?
    \item Never introduces a physical cause (a merger, an interaction, a lens, an AGN, a resonance, a starburst, a population state) that is not independently visible to you in the pixels right now -- a draft repeating literature-derived vocabulary is not itself confirmation that the cause is visible.
    \item Reports only what this specific image shows, describes only the apparent displayed color, and uses only frame-relative direction language.
    \item Ignores anything about identity, naming, discovery, or classification history -- for this object or any other.
    \item Is a single flowing paragraph of prose -- no headers, no bullet points, no markdown -- with length matched to how much is actually visible. An unresolved or visually simple target should get 2--4 sentences, not an exhaustive breakdown.
\end{itemize}

Before you answer, check your own response one more time: does it contain any name, designation, or identifying string, for this object or any other? If so, remove it before responding. This step is high-risk because you were shown the real name -- treat that as a reason for extra caution, not less.
\end{tcolorbox}

\subsection{Spectrum caption generation}

\begin{tcolorbox}[
    colback=gray!5!white,
    colframe=gray!75!black,
    fonttitle=\bfseries\small,
    fontupper=\footnotesize,
    title=Spectra Captioning Prompt,
    sharp corners,
    breakable,
    left=2mm, right=2mm, top=2mm, bottom=2mm
]
\ifdefined\setlist
  \setlist{nosep, leftmargin=1.3em, topsep=1pt, partopsep=0pt}
\fi
\setlength{\parskip}{2pt}

You are an expert astrophysicist writing a caption for an astronomical spectrum. The caption will be used to train a multimodal AI model that aligns 1D spectral data with natural language descriptions.

\medskip
You are provided with two sources of information about this object:
\begin{enumerate}
    \item \textbf{The Spectral Plot (Primary)}: An image of the 1D spectrum.
    \item \textbf{Literature Quotes (Secondary)}: A list of quotes from scientific papers that reference this object.
\end{enumerate}

\medskip
The object is located at a $\langle$\textbf{\texttt{redshift\_description}}$\rangle$ distance. State this distance regime plainly in your caption.

The plot includes labeled vertical dashed lines marking the expected observed positions of common rest-frame spectral features (redshifted by its distance). These include:
\begin{itemize}
    \item \textbf{Hydrogen Balmer series}: H$\alpha$, H$\beta$, H$\gamma$, H$\delta$ (and Ly$\alpha$ for high-$z$ objects)
    \item \textbf{Forbidden nebular emission}: [O~II]~$\lambda$3727, [O~III]~$\lambda\lambda$4959,5007, [O~I]~$\lambda$6300, [N~II]~$\lambda$6583, [S~II]~$\lambda\lambda$6716,6731
    \item \textbf{High-ionization/UV lines}: C~IV~$\lambda$1549, C~III]~$\lambda$1909, Mg~II~$\lambda$2799, He~II~$\lambda$4686
    \item \textbf{Stellar absorption features}: Ca~II H \& K, G-band, Mg~I b, Na~I D
\end{itemize}

A labeled line appearing on the plot does NOT mean the feature is detected---only report features where you can see a clear signal above or below the continuum.

\medskip
\noindent\textbf{OBJECT CLASSIFICATION (stated fact---do not hedge or re-derive)}\par\smallskip
State plainly, near the beginning of the caption, the object's classification: $\langle$\textbf{\texttt{class\_phrase}}$\rangle$ and subclass: $\langle$\textbf{\texttt{subclass\_phrase}}$\rangle$. You may weave these terms naturally into your prose, but they must be stated as established facts.

\medskip
\noindent\textbf{MEASURED EMISSION LINES (source of truth---do not detect independently)}\par\smallskip
The following emission lines have been measured in this spectrum with significant signal: $\langle$\texttt{desi\_lines}$\rangle$. Select the most salient 2--3 of these to mention, describing their visual appearance in the plot (e.g. narrow/broad, relative strength) rather than re-deriving whether they exist. Do not mention any emission or absorption line outside this list, and do not imply other lines are present. Do not include the numeric SNR values in your caption. Use the SNR values only internally to determine relative strengths and qualitative descriptions (e.g. `strong', `faint').

\noindent\textbf{THE ANALYSIS HIERARCHY}\par\smallskip
The plot is the \textbf{ABSOLUTE GROUND TRUTH} for what is observable. The quotes are only to be used to enrich the description of features that are already visible in the plot.

Follow this two-phase analysis before writing your caption:

\smallskip
\noindent\textbf{Phase 1: Analyze the Plot (What is visible?)}
\begin{itemize}
    \item What is the continuum shape (blue, red, flat, breaks)?
    \item Which emission or absorption lines are clearly visible? Are they broad or narrow?
    \item What is the general signal quality?
    \item Based \textit{only} on the plot, what is the rough classification (e.g., star-forming galaxy, quasar, early-type galaxy)?
\end{itemize}

\noindent\textbf{Phase 2: Filter and Enrich with Quotes (What details can we add?)}
\begin{itemize}
    \item Review the quotes and extract \textit{only} quantitative details or precise classifications (e.g., equivalent widths, velocity dispersions, line ratios, specific AGN types like ``Seyfert 1.5'') that correspond to the features you identified in Phase 1.
    \item \textbf{CRITICAL REJECTION}: Ignore ANY quotes that describe observations outside the wavelength range shown in the plot. Specifically, \textbf{REJECT} any information about:
    \begin{itemize}
        \item X-ray emission or absorption
        \item Radio, millimeter, or submillimeter observations (e.g., CO, HCN, megamasers)
        \item Far-infrared or mid-infrared features (e.g., silicate absorption, PAHs, [C~II]~158~$\mu$m)
    \end{itemize}
    \item \textbf{CONFLICT RESOLUTION}: If a quote claims a feature is strong (e.g., ``strong H$\alpha$''), but the plot shows no such feature, \textbf{TRUST THE PLOT} and do not mention it.
\end{itemize}

\medskip
\noindent\textbf{CRITICAL RULES}\par\smallskip
\begin{enumerate}
    \item \textbf{ONLY describe information that can be directly observed in the provided plot, enriched by relevant quotes.}
    \item \textbf{NEVER include any of the following, even if mentioned in the quotes:}
    \begin{itemize}
        \item Object names, catalog identifiers, or specific coordinates.
        \item Information from unobserved wavebands (X-ray, radio, far-IR/mid-IR).
        \item Morphological descriptions (size, shape, spiral arms, merger status).
        \item Distance or mass estimates beyond what is directly implied by the spectrum and redshift.
        \item The exact numeric redshift value (e.g., $z = 0.123$), use qualitative distance descriptions instead.
    \end{itemize}
    \item \textbf{Write in third person}, describing the spectrum directly (e.g., ``The spectrum displays strong narrow emission lines\dots'', ``This early-type galaxy spectrum is characterized by\dots''). Do not say ``the quotes state'' or ``the plot shows''.
    \item \textbf{Do NOT mention the object by name.} Refer to it by its spectral nature (e.g., ``this galaxy'', ``this quasar'', ``this spectrum'').
    \item If the spectrum is completely flat, uncalibrated, corrupted by noise, or contains no usable spectral signal, output exactly: \texttt{INSUFFICIENT\_SPECTRAL\_DATA}
\end{enumerate}

\medskip
\noindent\textbf{LITERATURE ENRICHMENT (qualitative only)}\par\smallskip
Include at most two qualitative details from the quotes below---e.g. a more precise subtype/classification, a noted peculiarity, or descriptive context---and only if consistent with the classification, plot, or measured lines above. Do not include quantitative figures (equivalent widths, velocity dispersions, column densities, line ratios, flux or mass values) from the quotes, even if available. If more than two qualifying qualitative details exist, choose the two most specific and best supported.

\begin{itemize}
    \item $\langle$\texttt{quote 1}$\rangle$
    \item $\langle$\texttt{quote 2}$\rangle$
    \item \dots
\end{itemize}

\medskip
\noindent\textbf{YOUR CAPTION}\par\smallskip
Write a concise, precise caption (2--4 sentences) that synthesizes the visual plot and the relevant, filtered quotes into a single cohesive description of the spectrum.

\end{tcolorbox}

\subsection{Light curve caption generation}

For each of the 987 BTS transients, Gemini 3.7 Flash received one ATCAT-aligned view of the accepted $g$- and $r$-band observations. We retained the interval from 30 days before the first positive detection with ${\rm S/N}>5$ through the last such detection, selected at most 243 rows using the encoder's fixed sampling rule, sorted them by time, and shifted the first retained observation to day zero. The request contained a deterministic summary, the complete retained table, a plot of the same rows, selected literature passages, and the authoritative broad class. The class was used only to condition the synthetic caption target
and was not supplied as an input to AstroBridge during training
or evaluation.

\begin{tcolorbox}[
    colback=gray!5!white,
    colframe=gray!75!black,
    fonttitle=\bfseries\small,
    fontupper=\footnotesize,
    title=Light curve Captioning Prompt,
    sharp corners,
    breakable,
    left=2mm, right=2mm, top=2mm, bottom=2mm
]
\setlength{\parskip}{2pt}

You are an expert time-domain astrophysicist. Write a concise scientific caption describing the optical light curve of one transient.

\medskip
The required broad class is:

\texttt{<target\_class>}\\
$\langle$\textbf{\texttt{target\_broad\_class}}$\rangle$\\
\texttt{</target\_class>}

\medskip
The class will be exactly one of \texttt{SN Ia}, \texttt{SN II}, or \texttt{SN Ibc}. It is an output constraint, not observational evidence. Do not mention that it was supplied, required, catalogued, or obtained from metadata. Do not state another class or a more specific subtype.

\medskip
The following deterministic summary was calculated only from the retained light curve rows below:

\texttt{<photometry\_summary>}\\
$\langle$\textbf{\texttt{photometry\_summary}}$\rangle$\\
\texttt{</photometry\_summary>}

\medskip
The complete retained light curve table is:

\texttt{<light\_curve>}\\
\texttt{TIME\_DAYS BAND FLUXCAL FLUXCAL\_ERR DIFF\_MAG DIFF\_MAG\_ERR}\\
$\langle$\textbf{\texttt{retained light curve rows}}$\rangle$\\
\texttt{</light\_curve>}

\medskip
\texttt{TIME\_DAYS} is measured from the first retained row. \texttt{FLUXCAL} and \texttt{FLUXCAL\_ERR} use the SNANA zero point 27.5. \texttt{FLUXCAL} is signed difference flux. \texttt{DIFF\_MAG} is
\[
27.5 - 2.5\log_{10}\!\left(\left|\texttt{FLUXCAL}\right|\right),
\]
and \texttt{DIFF\_MAG\_ERR} is its propagated uncertainty. The flux sign remains authoritative for whether a residual is positive or negative.

\medskip
An image of exactly the same retained rows is attached. Use it to inspect the overall morphology and use the table as the authority for exact measurements. The image has an inverted difference-magnitude axis, separates $g$ and $r$, shows uncertainty, and distinguishes positive and negative residuals. The summary, table, and image contain no independent measurements beyond one another.

\medskip
Additional literature passages are supplied below, unchanged from the established corpus:

\texttt{<private\_context>}\\
$\langle$\textbf{\texttt{selected literature passages}}$\rangle$\\
\texttt{</private\_context>}

\medskip
The private context is background only. It may guide which behavior to inspect in the retained light curve, but no output claim about a photometric value, time, duration, upper limit, band, light curve behavior, or sampling may rely on it. Every factual or quantitative light curve claim in the caption must be directly checkable from the retained table. Do not mention the private context or its sources.

Keep the caption focused on the retained light curve. Do not report an object or host name, coordinate, redshift, distance, absolute magnitude, spectrum, spectral feature, multiwavelength observation, progenitor, explosion mechanism, or event phase. Do not present any such information as the reason for the final classification.

\medskip
Use the following general guidance cautiously:
\begin{itemize}
    \item SN Ia light curves commonly show a relatively smooth rise and decline. The bluer band may fade faster after the brightest sampled region, while the redder band may show a slower decline or shoulder.
    \item SN II light curves may show a prolonged plateau, a sustained slow decline, or a longer-lasting evolution.
    \item SN Ibc light curves commonly show a comparatively rapid, single-peaked rise and decline without a prolonged plateau, but their photometric evolution can overlap substantially with SN Ia and SN II.
    \item These are population-level tendencies, not diagnostic rules. A missing feature is evidence only when the relevant phase was adequately sampled. Sparse or gapped coverage means that the feature is unknown, not absent.
\end{itemize}

\medskip
Describe only supported behavior such as the retained time span, sampled rise or decline, changes in slope, shoulders or multiple sampled maxima, band-dependent evolution, gaps, negative residuals, and measurement uncertainty. Refer to the ``brightest sampled point'' rather than the true peak unless the rows clearly sample a turnover on both sides. Compare bands as colour information only when their measurements are sufficiently close in time.

\medskip
Return only a self-contained caption of 2--4 sentences. Use the first sentences to describe the observed photometric evolution without stating a class. In the final sentence, connect only the photometric features already described to the required broad class and conclude using wording of the form ``\dots is consistent with the [class] class.'' Never use ``confirms'', ``proves'', ``establishes'', or similarly strong wording. Do not invent, exaggerate, or selectively omit observations to make the required class appear better supported. If the features are not individually diagnostic, state this explicitly.

\medskip
If the supplied light curve evidence does not support a meaningful object-specific description, return exactly:

\begin{center}
\texttt{INSUFFICIENT\_PHOTOMETRIC\_DATA}
\end{center}

\end{tcolorbox}

\subsection{Model training}

During training, we randomly sampled one system prompt and one user instruction per observation. In the user instructions, the \texttt{\{modalities\}} placeholder was dynamically replaced with the specific observation type being passed.

\begin{tcolorbox}[
    colback=gray!5!white,
    colframe=gray!75!black,
    fonttitle=\bfseries\small,
    fontupper=\small,
    title=Training Prompt Variants,
    sharp corners,
    breakable,
    left=2mm, right=2mm, top=2mm, bottom=2mm
]
\textbf{System Prompt Variants (Randomly Sampled):}
\begin{itemize}
    \item ``You are an astronomy assistant. Describe only what the provided observation supports.''
    \item ``You are an expert astronomy assistant. Describe only what is supported by the provided observation. Do not infer properties that cannot be reliably determined from it; when evidence is insufficient, say so rather than guessing.''
    \item ``You are an astronomy assistant. Base your description strictly on the observation given; do not infer properties it cannot show.''
    \item ``You are an astronomy assistant. Report what the observation shows, and state the limits of what it can establish.''
\end{itemize}

\vspace{0.8em}
\textbf{User Instruction Variants (Randomly Sampled):}
\begin{itemize}
    \item ``Describe the scientifically supported properties of this observation.''
    \item ``Summarize the main features present in this observation.''
    \item ``What can be reliably inferred from this observation?''
    \item ``Provide a concise astronomical caption grounded only in this observation.''
    \item ``Describe this observation, including relevant physical properties where they are supported.''
    \item ``Describe the object shown, using only \texttt{\{modalities\}}.''
    \item ``Based on \texttt{\{modalities\}} alone, characterise this object.''
    \item ``What does this \texttt{\{modalities\}} show?''
    \item ``Write a short caption for this \texttt{\{modalities\}}.''
    \item ``Summarise what \texttt{\{modalities\}} reveals about this object.''
\end{itemize}
\end{tcolorbox}

\subsection{Evaluation}

As a sample, we provide all the prompts used for the spectra-based evaluations. We provide the four prompts used by Gemini to extract physical properties from captions, followed by the three caption generation prompts. Unlike AstroBridge, the baseline models are not directly trained to extract physical information from spectra, so we provide them with more detailed prompts to give them the best chance of success and make the evaluation as fair as possible.

\begin{tcolorbox}[
    colback=gray!5!white,
    colframe=gray!75!black,
    fonttitle=\bfseries\small,
    fontupper=\small,
    title=Distance Classification Evaluation Prompt,
    sharp corners,
    breakable,
    left=2.5mm, right=2.5mm, top=2.5mm, bottom=2.5mm
]
\ifdefined\setlist
  \setlist{nosep, leftmargin=1.5em, topsep=2pt, partopsep=0pt}
\fi
\setlength{\parskip}{2.5pt}

You are an expert astrophysicist. You will be provided with a scientific description of an astronomical spectrum.
Based on the description of the spectrum, classify the distance of the observed astronomical object into one of the following categories.

\medskip
\noindent\textbf{Allowed categories:}
\begin{itemize}
    \item[\textbf{A:}] Very Close ($z < 0.1$)
    \item[\textbf{B:}] Close ($0.1 < z < 0.5$)
    \item[\textbf{C:}] Far ($z > 0.5$)
\end{itemize}

\medskip
\noindent\textbf{Spectrum Description:}\\
\texttt{"""}\\
$\langle$\textbf{\texttt{caption}}$\rangle$\\
\texttt{"""}

\medskip
Think step-by-step, but you \textbf{MUST} conclude your response with the exact format:\par\smallskip
\hspace*{1.5em}\texttt{\textbf{FINAL ANSWER: [Letter]}}
\end{tcolorbox}

\begin{tcolorbox}[
    colback=gray!5!white,
    colframe=gray!75!black,
    fonttitle=\bfseries\small,
    fontupper=\small,
    title=Emission-Line Extraction Evaluation Prompt,
    sharp corners,
    breakable,
    left=2.5mm, right=2.5mm, top=2.5mm, bottom=2.5mm
]
\ifdefined\setlist
  \setlist{nosep, leftmargin=1.5em, topsep=2pt, partopsep=0pt}
\fi
\setlength{\parskip}{2.5pt}

You are an expert astrophysicist. You will be provided with a scientific description of an astronomical spectrum.
Based on the description of the spectrum, evaluate the following candidate checklist of emission lines and determine which ones are present, detected, or evidenced in the spectrum.

\medskip
\noindent\textbf{Allowed candidate lines:}
\begin{itemize}
    \item[\textbf{A:}] $\langle$\texttt{candidate\_line\_1}$\rangle$
    \item[\textbf{B:}] $\langle$\texttt{candidate\_line\_2}$\rangle$
    \item[\phantom{\textbf{A:}}] \hspace*{-0.6em}\dots
    \item[\textbf{\textit{N}:}] $\langle$\texttt{candidate\_line\_n}$\rangle$
\end{itemize}

\medskip
\noindent\textbf{Spectrum Description:}\\
\texttt{"""}\\
$\langle$\textbf{\texttt{caption}}$\rangle$\\
\texttt{"""}

\medskip
Think step-by-step, but you \textbf{MUST} conclude your response with the exact format:\par\smallskip
\hspace*{1.5em}\texttt{\textbf{FINAL ANSWER: [Letter(s) separated by comma, e.g. A, B]}}\par\smallskip
If none of the candidate emission lines are present or mentioned, write:\par\smallskip
\hspace*{1.5em}\texttt{\textbf{FINAL ANSWER: NONE}}
\end{tcolorbox}

\begin{tcolorbox}[
    colback=gray!5!white,
    colframe=gray!75!black,
    fonttitle=\bfseries\small,
    fontupper=\small,
    title=Galaxy vs. Quasar Evaluation Prompt,
    sharp corners,
    breakable,
    left=2.5mm, right=2.5mm, top=2.5mm, bottom=2.5mm
]
\ifdefined\setlist
  \setlist{nosep, leftmargin=1.5em, topsep=2pt, partopsep=0pt}
\fi
\setlength{\parskip}{2.5pt}

You are an expert astrophysicist. You will be provided with a scientific description of an astronomical spectrum.
Based on the description of the spectrum, classify the astronomical source class into one of the following categories.

\medskip
\noindent\textbf{Allowed categories:}
\begin{itemize}
    \item[\textbf{A:}] Galaxy
    \item[\textbf{B:}] Quasar
\end{itemize}

\medskip
\noindent\textbf{Spectrum Description:}\\
\texttt{"""}\\
$\langle$\textbf{\texttt{caption}}$\rangle$\\
\texttt{"""}

\medskip
Think step-by-step, but you \textbf{MUST} conclude your response with the exact format:\par\smallskip
\hspace*{1.5em}\texttt{\textbf{FINAL ANSWER: [Letter]}}
\end{tcolorbox}

\begin{tcolorbox}[
    colback=gray!5!white,
    colframe=gray!75!black,
    fonttitle=\bfseries\small,
    fontupper=\small,
    title=Source Type Classification Evaluation Prompt,
    sharp corners,
    breakable,
    left=2.5mm, right=2.5mm, top=2.5mm, bottom=2.5mm
]
\ifdefined\setlist
  \setlist{nosep, leftmargin=1.5em, topsep=2pt, partopsep=0pt}
\fi
\setlength{\parskip}{2.5pt}

You are an expert astrophysicist. You will be provided with a scientific description of an astronomical spectrum.
Based on the description of the spectrum, classify the astronomical source into one of the following categories.

\medskip
\noindent\textbf{Allowed categories:}
\begin{itemize}
    \item[\textbf{A:}] AGN
    \item[\textbf{B:}] Starburst
    \item[\textbf{C:}] Starforming
    \item[\textbf{D:}] Broadline
\end{itemize}

\medskip
\noindent\textbf{Spectrum Description:}\\
\texttt{"""}\\
$\langle$\textbf{\texttt{caption}}$\rangle$\\
\texttt{"""}

\medskip
Think step-by-step, but you \textbf{MUST} conclude your response with the exact format:\par\smallskip
\hspace*{1.5em}\texttt{\textbf{FINAL ANSWER: [Letter]}}
\end{tcolorbox}

\begin{tcolorbox}[
    colback=gray!5!white,
    colframe=gray!75!black,
    fonttitle=\bfseries\small,
    fontupper=\small,
    title=AstroBridge Caption Generation Prompt,
    sharp corners,
    breakable,
    left=2.5mm, right=2.5mm, top=2mm, bottom=2mm
]
\setlength{\parskip}{2pt}

\noindent\textbf{System Prompt:}\\
You are an astronomy assistant. Describe only what the provided observation supports.

\medskip
\noindent\textbf{User Input:}\\
\texttt{<observation>}$\langle$\textbf{\texttt{Spectrum Prefix (64 tokens)}}$\rangle$\texttt{</observation>}\\
Describe this observation, including relevant physical properties where they are supported.
\end{tcolorbox}

\begin{tcolorbox}[
    colback=gray!5!white,
    colframe=gray!75!black,
    fonttitle=\bfseries\small,
    fontupper=\small,
    title=Vision Baseline Caption Generation Prompt (Plot),
    sharp corners,
    breakable,
    left=2.5mm, right=2.5mm, top=2mm, bottom=2mm
]
\setlength{\parskip}{2pt}

\noindent\textbf{System Prompt:}\\
You are an expert astronomer analyzing spectral data. After any internal reasoning, you MUST output your final caption on a new line beginning with ``CAPTION:'' followed by 2--4 concise sentences.

\medskip
\noindent\textbf{User Input:}\\
$\langle$\textbf{\texttt{Rendered Spectrum Plot Image}}$\rangle$\par\smallskip
Provide a concise astronomical caption (2--4 sentences) characterizing this spectrum observation. Summarize the continuum shape, prominent emission or absorption lines (e.g.~Balmer lines, [O~III]), and the physical nature of the object. Do not describe the plot layout, axes, colors, or image elements. Do not include step-by-step reasoning in the final caption. Begin your final caption on a new line with ``CAPTION:''.
\end{tcolorbox}

\begin{tcolorbox}[
    colback=gray!5!white,
    colframe=gray!75!black,
    fonttitle=\bfseries\small,
    fontupper=\small,
    title=Text Baseline Caption Generation Prompt (Text),
    sharp corners,
    breakable,
    left=2.5mm, right=2.5mm, top=2mm, bottom=2mm
]
\setlength{\parskip}{2pt}

\noindent\textbf{System Prompt:}\\
You are an expert astronomer analyzing spectral data. After any internal reasoning, you MUST output your final caption on a new line beginning with ``CAPTION:'' followed by 2--4 concise sentences.

\medskip
\noindent\textbf{User Input:}\\
Provide a concise astronomical caption (2--4 sentences) characterizing this spectrum observation. Summarize the continuum shape, prominent emission or absorption lines (e.g.~Balmer lines, [O~III]), and the physical nature of the object. Do not include step-by-step reasoning in the final caption. Begin your final caption on a new line with ``CAPTION:''.

\medskip
\noindent\textbf{Appended Spectrum Text:}\\
\texttt{Wavelength [Angstrom]: [}$\lambda_1, \lambda_2, \dots, \lambda_{100}$\texttt{]}\\
\texttt{Flux: [}$f_1, f_2, \dots, f_{100}$\texttt{]}
\end{tcolorbox}

\section{Light Curve Caption Analysis}
\label{app:lightcurve-caption-analysis}

These diagnostics were designed after inspecting the benchmark results and are
explanatory. We use Gemini 3.7 Flash to extract explicit
duration, peak, and gap claims from the 266 AstroBridge captions, then we deterministically performed all comparisons.
Duration was measured from the first to last retained observation, peaks from
the maximum signed \texttt{FLUXCAL} in each band, and gaps from consecutive
observations. A fixed seed-42 sample of 20 captions (7 SN Ia, 7 SN II, and
6 SN Ibc) confirmed correct extraction of all 20 duration, 39 peak, and
12 gap claims, with none missed in these categories.
\medskip
\begin{center}
\begin{minipage}{0.9\linewidth}
\refstepcounter{table}
\textbf{Table \thetable:} Absolute errors in explicit numerical caption
claims. Counts are scorable claims.
\label{tab:caption-numeric-errors}
\vspace{0.5em}

\centering
\small
\begin{tabular}{lrrrr}
\toprule
Claim & Count & Mean & Median & 90th percentile \\
\midrule
Duration (days)      & 266 & 29.10 & 16.09 & 44.93 \\
Peak time (days)     & 517 &  9.92 &  4.89 & 14.00 \\
Peak magnitude (mag) & 517 &  0.456 & 0.363 & 0.917 \\
Gap endpoint (days)  & 121 &  9.25 &  6.05 &  9.95 \\
\bottomrule
\end{tabular}
\end{minipage}
\end{center}
\medskip
\medskip
\begin{center}
\begin{minipage}{\linewidth}
\begin{minipage}[t]{0.32\linewidth}
  \centering
  \includegraphics[width=\linewidth]{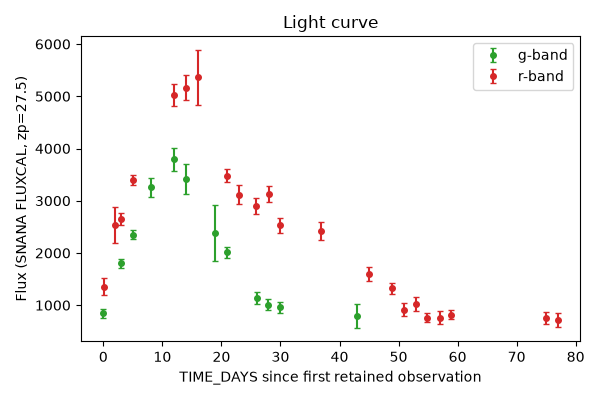}

  \small
  \textbf{SN2020aatr: Ia $\rightarrow$ Ia.}\\
  Reported duration: 77 days; error: 0.05 days.
\end{minipage}
\hfill
\begin{minipage}[t]{0.32\linewidth}
  \centering
  \includegraphics[width=\linewidth]{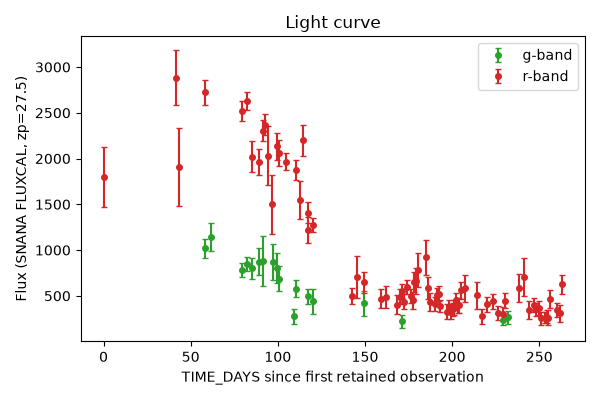}

  \small
  \textbf{SN2020ej: II $\rightarrow$ II.}\\
  Reported duration: 1100 days; retained span: 263 days.
\end{minipage}
\hfill
\begin{minipage}[t]{0.32\linewidth}
  \centering
  \includegraphics[width=\linewidth]{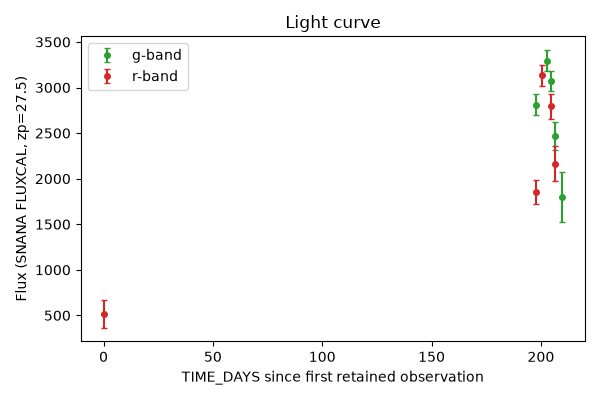}

  \small
  \textbf{SN2021sje: Ibc $\rightarrow$ Ia.}\\
  Reported duration: 34 days; retained span: 210 days.
\end{minipage}

\medskip
\refstepcounter{figure}
\small\textbf{Figure \thefigure:}
A well-grounded caption, a correct classification with a severe numerical
error, and a failure in both numerical description and class.
\label{fig:lightcurve-caption-examples}
\end{minipage}
\end{center}
\medskip

We offer several plausible explanations for these issues. Firstly, the frozen ATCAT encoder, pretrained on simulated light curves, transfers imperfectly to real ZTF observations. Secondly, the caption-tuned 9B decoder seems to learn common descriptive patterns more readily than exact numerical operations over compressed representations. In addition, non-thinking generation and the absence of calculation tools may further encourage fluent, template-like captions. Future work can test these hypotheses through controlled ablations of real-data encoder adaptation, decoder capacity, reasoning mode, and numerical tool use.

\section{Full Spectra Results}
\label{app:full-spectra-results}

We evaluated spectra understanding on four held-out datasets sampled from SDSS DR17 and DESI EDR, with all training-set observations strictly excluded. Ground-truth labels for the three classification tasks were obtained via NOIRLab DataLab. The class-balanced evaluation sets are constructed as follows: Quasar vs. Galaxy uses 200 samples equally split across classes and surveys (SDSS/DESI); Redshift uses 300 samples evenly divided across three bins  ($z < 0.1$, $0.1 < z < 0.5$, $z>0.5$)  for both surveys, Source Type comprises 200 SDSS-only samples balanced across four classes (star-forming, starburst, AGN, broadline), and Emission Lines utilizes 200 DESI-only samples balanced across four detection regimes.

To test the robustness of the model across varying signal strengths, the evaluator LLM is given 3-4 candidate lines and must identify which are present, if any. The 200-sample test set is divided evenly into four regimes: (1) Absent: no queried lines present (measured by the \% of observations correctly reporting no lines); (2) All Strong: all lines clearly visible (measured by recall); (3) Mixed (Strong): 1-2 lines present (SNR > 5), others absent (measured by Jaccard index); and (4) Mixed (Faint): 1-2 faint lines present (SNR > 3), others absent (Jaccard index). The Jaccard index is computed per observation and then averaged. When the ground-truth set of candidate lines is empty, the score
is 1 if the predicted set is also empty and 0 otherwise. The value reported in Table ~\ref{tab:results} is the average of the four regimes.

To make the comparison fair, we use inference budgets and fallback protocols throughout the pipeline. During the initial caption generation phase, all evaluated models were allowed to generate intermediate reasoning steps with a maximum budget of 2048 tokens. If a model exhausted this budget without outputting the required \texttt{CAPTION:} delimiter, we appended the string \verb|\n\nFinal answer below.\nCAPTION:| and forced the model to continue generation with an additional 256 tokens. Similarly, during the downstream property extraction phase, the frozen Gemini evaluator operated under a 1024-token budget. If it failed to produce a parseable response ending with the \texttt{FINAL ANSWER:} indicator, we supplied the follow-up prompt ``Please continue your previous response and output only the final answer starting with the indicator \texttt{FINAL ANSWER:}'' and granted an additional 128 tokens to complete the extraction.

The results across all tasks and regimes are presented in Table~\ref{tab:spectra-results-full}. This table also includes two additional rows for models that were given the spectrum in text format, which is a subsampled list of wavelength-flux pairs. Overall, AstroBridge significantly outperforms both the plot-rendered and text-based baselines across all classification tasks and the positive emission-line regimes. The only exception is the fully absent emission-line regime, where AstroBridge gets the second-lowest score. One possible explanation is that the baselines are more conservative and tend to predict no lines when uncertain, while AstroBridge is more aggressive in identifying lines. 


\begin{table*}[t]
\centering
\small
\caption{Spectra evaluation. Classification columns report accuracy (\%) on class-balanced test sets.Emission-line columns report the metric in parentheses. Baselines receive the spectrum as a rendered plot (Plot) or subsampled wavelength--flux list (Text); AstroBridge receives the raw 1-D spectrum through its encoder. Bold = best, underline = second best.}
\vspace{0.5em}
\label{tab:spectra-results-full}

\resizebox{\linewidth}{!}{%
\begin{tabular}{l ccc c cccc}
\toprule
& \multicolumn{3}{c}{Classification (Accuracy \%)} & & \multicolumn{4}{c}{Emission-line extraction} \\
\cmidrule(lr){2-4} \cmidrule(lr){6-9}
Model
& \shortstack{QSO vs.\\Galaxy}
& \shortstack{Source\\type}
& \shortstack{Redshift\\class}
&
& \shortstack{All Absent \\ (\% None)}
& \shortstack{All Present\\(Recall)}
& \shortstack{Mixed\\(Jaccard)}
& \shortstack{Mixed (Hard)\\(Jaccard)} \\
\midrule
Qwen 3.5 9B (Plot)       & 69.0 {\scriptsize $\pm$ 6.5} & \underline{41.5} {\scriptsize $\pm$ 7.0} & 42.3 {\scriptsize $\pm$ 5.7} && 92.0 {\scriptsize $\pm$ 7.0} & 14.0 {\scriptsize $\pm$ 7.5} & 5.0 {\scriptsize $\pm$ 4.3} & 4.0 {\scriptsize $\pm$ 4.5} \\
Gemma 4 12B (Text)       & 50.0 {\scriptsize $\pm$ 7.0} & 26.0 {\scriptsize $\pm$ 6.0} & 33.3 {\scriptsize $\pm$ 5.3} && 30.0 {\scriptsize $\pm$ 12.0} & 21.7 {\scriptsize $\pm$ 7.2} & \underline{19.0} {\scriptsize $\pm$ 8.4} & \underline{17.3} {\scriptsize $\pm$ 7.3} \\
Gemma 4 12B (Plot)       & 61.5 {\scriptsize $\pm$ 6.7} & 36.0 {\scriptsize $\pm$ 6.5} & 35.3 {\scriptsize $\pm$ 5.3} && 94.0 {\scriptsize $\pm$ 7.0} & 13.7 {\scriptsize $\pm$ 6.1} & 12.0 {\scriptsize $\pm$ 7.5} & 1.0 {\scriptsize $\pm$ 1.5} \\
Qwen 3.8 27B (Plot)      & \underline{73.0} {\scriptsize $\pm$ 6.0} & 41.0 {\scriptsize $\pm$ 6.8} & 35.7 {\scriptsize $\pm$ 5.5} && \underline{96.0} {\scriptsize $\pm$ 5.0} & \underline{27.3} {\scriptsize $\pm$ 9.3} & 6.7 {\scriptsize $\pm$ 5.3} & 4.0 {\scriptsize $\pm$ 4.5} \\
Muse Glimmer 30B (Plot)  & 59.0 {\scriptsize $\pm$ 7.0} & 38.5 {\scriptsize $\pm$ 6.8} & \underline{45.7} {\scriptsize $\pm$ 5.7} && \textbf{100.0} {\scriptsize $\pm$ 0.0} & 26.5 {\scriptsize $\pm$ 9.0} & 9.0 {\scriptsize $\pm$ 6.5} & 1.0 {\scriptsize $\pm$ 1.5} \\
Gemma 4 31B (Text)       & 51.0 {\scriptsize $\pm$ 7.0} & 25.5 {\scriptsize $\pm$ 6.2} & 31.3 {\scriptsize $\pm$ 5.3} && 66.0 {\scriptsize $\pm$ 13.0} & 10.7 {\scriptsize $\pm$ 5.3} & 16.0 {\scriptsize $\pm$ 8.2} & 10.0 {\scriptsize $\pm$ 6.0} \\
Gemma 4 31B (Plot)       & 51.0 {\scriptsize $\pm$ 7.0} & 32.0 {\scriptsize $\pm$ 6.2} & 40.0 {\scriptsize $\pm$ 5.5} && \underline{96.0} {\scriptsize $\pm$ 5.0} & 11.0 {\scriptsize $\pm$ 6.0} & 4.0 {\scriptsize $\pm$ 3.5} & 1.3 {\scriptsize $\pm$ 1.7} \\
\textbf{AstroBridge}     & \textbf{91.5} {\scriptsize $\pm$ 3.8} & \textbf{56.5} {\scriptsize $\pm$ 7.0} & \textbf{90.3} {\scriptsize $\pm$ 3.3} && 42.0 {\scriptsize $\pm$ 14.0} & \textbf{66.8} {\scriptsize $\pm$ 8.8} & \textbf{57.7} {\scriptsize $\pm$ 11.3} & \textbf{34.2} {\scriptsize $\pm$ 10.1} \\
\bottomrule
\end{tabular}%
}
\end{table*}

\section{Full Light Curve Results}
\label{app:full-lightcurve-results}

We evaluated transient classification capabilities of several models on all held-out YSE objects: AstroBridge receives light curves through its frozen ATCAT encoder and trained fusion network, while base Qwen3.5 9B and Gemma 4 12B/31B receive either raw observations as a table or a plot. 
We also want to see how much information Q-former representations manage to retain, so we freeze the trained model and extract its pre-adapter $64\times384$ Q-former representation for each light curve. We then fit a linear classifier and a
one-layer transformer to these representations.
Training settings are selected by five-fold stratified cross-validation on the 987 BTS objects; the final heads are fitted on all 987 objects and evaluated once on YSE.
\begin{table*}[t]
\centering
\small
\caption{Light curve classification on the object-disjoint YSE test set.
Accuracy intervals are 95\% class-stratified bootstrap intervals from 10,000
resamples of the saved predictions. Class columns report recall.}
\label{tab:lightcurve-results-full}
\resizebox{\textwidth}{!}{%
\begin{tabular}{lccccccc}
\toprule
Model&
Accuracy [95\% CI] &
Macro F1 (\%)&
Balanced acc. (\%)&
Parse rate (\%) &
SN Ia ($n=180$) (\%) &
SN II ($n=71$) (\%) &
SN Ibc ($n=15$) (\%) \\
\midrule
\textbf{AstroBridge}
& \textbf{77.07} [72.93, 80.83]
& 52.25 & 50.23 & 100.00
& \textbf{93.33} & 50.70 & 6.67 \\

Qwen3.5-9B (table)
& 34.96 [29.70, 40.23]
& 25.80 & 32.05 & 77.82
& 46.67 & 2.82 & 46.67 \\

Qwen3.5-9B (plot)
& 58.65 [54.14, 62.78]
& 33.22 & 38.17 & 98.50
& 82.22 & 5.63 & 26.67 \\

Gemma 4 12B (table)
& 48.50 [42.68, 54.14]
& 36.29 & 38.29 & 100.00
& 57.22 & 30.99 & 26.67 \\

Gemma 4 12B (plot)
& 60.15 [56.39, 63.91]
& 34.84 & 36.59 & 90.23
& 85.56 & 4.23 & 20.00 \\

\textbf{Gemma 4 31B (table)}
& 69.17 [63.91, 74.44]
& \textbf{52.67} & 54.16 & 100.00
& 76.67 & \textbf{59.15} & 26.67 \\

Gemma 4 31B (plot)
& 69.55 [64.66, 74.44]
& 43.59 & 44.78 & 100.00
& 82.22 & 52.11 & 0.00 \\

\textbf{Linear probe}
& 56.77 [50.75, 62.78]
& 48.03 & \textbf{55.68} & --
& 60.56 & 46.48 & 60.00 \\

Transformer head
& 48.87 [42.86, 54.89]
& 43.25 & 51.84 & --
& 52.22 & 36.62 & \textbf{66.67} \\
\bottomrule
\end{tabular}}
\end{table*}

AstroBridge obtains the highest overall accuracy, exceeding the 67.67\% always-SN-Ia baseline, but its performance is concentrated in SN Ia: it
recovers 168/180 SN Ia objects and only 1/15 SN Ibc objects. Gemma 31B with the numerical table is the strongest generative route by macro F1, while the
linear probe has the highest balanced accuracy and recovers 9/15 SN Ibc objects. The probe therefore recovers minority-class information from the
frozen AstroBridge representation that the caption-trained decoder rarely expresses. Finally, although AstroBridge's captions were fluent, their precise numerical claims
were often unsupported by the raw observations. An additional audit found median absolute errors of 16.09 days for duration and 4.89 days
for peak time; Appendix~\ref{app:lightcurve-caption-analysis} gives the full
procedure, results, and some examples.

\subsection{Probing pipeline stages}
\label{app:lc-probing}

We also checked whether the Q-former or the projector lose information that the encoder already has. We average the representation at four points in the pipeline into a single vector and train a logistic regression classifier on it. The four points are:
(0) the raw ATCAT encoder output;
(1) our trained Q-former output, before the adapter;
(2) our trained Q-former output after the projector into the LLM embedding space.
We weight the classes equally during training, which makes the classifier predict rare classes like SN Ibc more often. Averaging removes the order of the tokens, so these results may understate what the representations hold.

We use 5-fold cross-validation on the YSE test set ($n=266$), since we have no other unseen light curves. This set is heavily imbalanced. This setup differs from the BTS-trained probe in Table~\ref{tab:lightcurve-results-full}, so the numbers cannot be compared directly. We compare stages with McNemar's test and bootstrap 95\% confidence intervals.

\begin{table}[ht]
\centering
\small
\caption{Probe results at each pipeline stage on YSE, from 5-fold cross-validation.
SN Ibc columns are precision (P), recall (R) and F1 on the 15 SN Ibc objects (5.6\% of the set).}
\label{tab:lc-probing}
\resizebox{\columnwidth}{!}{%
\begin{tabular}{lccccccc}
\toprule
Probe input & Dim & Acc. & Macro-F1 & \multicolumn{3}{c}{SN Ibc} \\
\cmidrule(lr){5-7}
& & & & P & R & F1 \\
\midrule
Raw ATCAT encoder               & 384  & 0.696 & 0.559 & 0.13 & 0.47 & 0.20 \\
Trained Q-former (pre-adapter)  & 384  & 0.711 & 0.587 & 0.17 & 0.60 & 0.27 \\
Trained Q-former (post-adapter) & 4096 & 0.718 & 0.583 & 0.18 & 0.60 & 0.28 \\
\bottomrule
\end{tabular}}
\end{table}

All three probes score about the same. None of the six pairwise differences is significant, so we find no sign that the Q-former or projector lose information from the encoder. 

SN Ibc precision is low at every stage (0.13 to 0.18), only 2 to 3 times the 5.6\% base rate. Recall is higher (0.47 to 0.60), but this comes from predicting SN Ibc often, and most of those predictions are SN Ia or SN II objects. The raw encoder, which involves no Q-former or LLM, shows the same problem, so the weakness starts at or before the encoder and not only in the language model. With only 15 SN Ibc objects each number is noisy, so the consistency across all four stages matters more than any single value. The BTS-trained linear probe and transformer head in Table~\ref{tab:lightcurve-results-full} find more SN Ibc objects (9/15 and 10/15), but they miss many more SN Ia (recall 60.56\% and 52.22\%, versus 93.33\% for AstroBridge).

\section{Full Image Results}
\label{app:full-image-results}

\begin{table}[ht]
\centering
\caption{Galaxy10~\citep{leung2024galaxy10} morphology classification: base LLM vs.\ AstroBridge (Galaxy10-AION, $n=150$, seed 0). `Shuffled' re-runs the identical objects under a deranged digit$\to$class code mapping, as a diagnostic against a suspected digit-token bias.}
\label{tab:galaxy10-eval}
\begin{tabular}{lcccc}
\toprule
& \multicolumn{2}{c}{Default codes} & \multicolumn{2}{c}{Shuffled codes} \\
\cmidrule(lr){2-3} \cmidrule(lr){4-5}
Metric & Base & Equipped & Base & Equipped \\
\midrule
Hard accuracy      & 0.513 & 0.313 & 0.507 & 0.260 \\
Macro-F1           & 0.372 & 0.154 & 0.355 & 0.135 \\
Group score         & 0.627 & 0.437 & 0.633 & 0.320 \\
Debiased soft score & --    & --    & 0.644 & 0.389 \\
\bottomrule
\end{tabular}
\end{table}

\begin{table}[ht]
\centering
\caption{Galaxy10-AION classification-from-caption accuracy across all captioning
approaches evaluated in this investigation. All accuracies on the 794-object held-out test split. Accuracy scores are computed on successful API calls, therefore a request that returns an unparseable file does not contribute to the final score.}
\label{tab:classification-master}
\begin{tabular}{llc}
\toprule
\textbf{Method}  & \textbf{Judge} & \textbf{Accuracy} \\
\midrule
Native Qwen   & Claude & 43.32\% \\
Native Qwen   & Gemini & 42.68\% \\
\midrule
\textbf{AstroBridge} & Gemini & \textbf{38.04\%} \\
\textbf{AstroBridge}  & Claude & \textbf{37.15\%} \\
\bottomrule
\end{tabular}
\end{table}

Across every metric and both digit mappings, AstroBridge performs worse compared to the unmodified base LLM. The fusion-stack/LoRA pipeline does not improve (rather degrades). The model struggles to/refuses to output classes corresponding to [0, 3, 8, 9], rather persists regardless of which class is attached to them. There is also true visual confusion on fine-grained sub-classes (Unbarred Tight/Loose Spiral, In-between Round Smooth) which is noticeable irrespective of which class index is used to represent it.

We note that classification accuracy from captioning may reward hedging from more verbose models like native Qwen, and penalize a concise captioner like AstroBridge, which writes shorter and more committed text. Pairwise judges (Claude, Gemini, GPT Sol) consistently rate AstroBridge's captions as higher quality, with better feature descriptions and less prone to hallucinations. In AstroBridge, our captioning prompts (App. \ref{app:image-caption-prompts}) avoid expert interpretations by design, opting for descriptions that conform to observable image features. While this increases caption quality, it may be the cause of poor  downstream accuracy on GZ10 image classification.

\subsection{Probing pipeline stages}
\label{app:img-probing}

We run the same check as in App.~\ref{app:lc-probing}, using AION as the raw encoder. The untrained Q-former result is the mean of 3 seeds. We use a fixed Galaxy10 split ($n_{\text{train}}=2003$, $n_{\text{test}}=498$). This differs from the 794 and 150 object splits used above, so the numbers cannot be compared directly.

\begin{table}[ht]
\centering
\small
\caption{Galaxy10 probe accuracy and macro-F1 at each pipeline stage.}
\label{tab:img-probing}
\begin{tabular}{lccc}
\toprule
Probe input & Dim & Acc. & Macro-F1 \\
\midrule
Raw AION encoder                & 768  & 0.661 & 0.661 \\
Untrained Q-former              & 384  & 0.631 & 0.637 \\
Trained Q-former (pre-adapter)  & 384  & 0.649 & 0.664 \\
Trained Q-former (post-adapter) & 4096 & 0.647 & 0.657 \\
\bottomrule
\end{tabular}
\end{table}

All four probes score about the same, and none of the six pairwise differences is significant. As with light curves, we find no sign that the Q-former or projector lose information from the encoder.


\end{document}